\documentclass[twocolumn,epjc3,preprint]{svjour3new}
\usepackage{color}
\usepackage{float}
\usepackage[colorlinks]{hyperref}
\usepackage{subfigure}
\usepackage{xspace}
\usepackage{mathtools}
\usepackage[utf8]{inputenc}
\usepackage{mathtools, cuted}
\usepackage{lipsum, color}
\usepackage{amsmath}
\usepackage{slashbox}
\usepackage{graphicx}
\usepackage{booktabs,multirow}

\smartqed  
\RequirePackage{graphicx}
\journalname{Eur. Phys. J. C}
\definecolor{darkred}{rgb}{0.8,0.0,0.0}

\newcommand{\nn}{\nonumber}
\newcommand{\lt}{\left}
\newcommand{\rt}{\right}
\newcommand{\fig}[1]{Fig.~\ref{#1}}
\newcommand{\eq}[1]{Eq.~(\ref{#1})}

\begin{document}

\title{Higgs portal to dark matter and \boldmath $B\to K^{(*)}$\unboldmath\
  decays}

\titlerunning{Higgs portal to dark matter and $B\to K^{(*)}$
  }        

\author{Aliaksei Kachanovich\thanksref{e1,addr1} \and
  Ulrich Nierste\thanksref{e2,addr1},
          and
  Ivan Ni\v sand\v zi\'c\thanksref{e3,addr1}
}

\thankstext{e1}{e-mail: aliaksei.kachanovich@kit.edu}
\thankstext{e2}{e-mail: ulrich.nierste@kit.edu}
\thankstext{e3}{e-mail: ivan.nisandzic@kit.edu}


\institute{Institute for Theoretical Particle Physics (TTP),
  Karlsruhe Institute of Technology (KIT),
  Wolfgang-Gaede-Stra\ss e 1, 
        76131 Karlsruhe, Germany\label{addr1}} 

\date{}

\maketitle

\begin{abstract}
We consider a Higgs portal model in which the 125-GeV Higgs boson mixes
with a light singlet mediator $h_2$ coupling to particles of a Dark
Sector and study potential $b\to s h_2$ decays in the Belle II
experiment.  Multiplying the gauge-dependent off-shell Stan\-dard-Model
$b$-$s$-Higgs vertex with the sine of the Higgs mixing angle does not
give the correct $b$-$s$-$h_2$ vertex. We clarify this issue by
calculating the $b$-$s$-$h_2$ vertex in an arbitrary $R_\xi$ gauge and
demonstrate how the $\xi$ dependence cancels from physical decay rates
involving an on-shell or off-shell $h_2$. Then we revisit the $b\to s
h_2$ phenomenology and point out that a simultaneous study of $B\to K^*
h_2$ and $B\to K h_2$ helps to discriminate between the Higgs portal and
alternative models of the Dark Sector. We further advocate for the use
of the $h_2$ lifetime information contained in displaced-vertex data
with $h_2$ decaying back to Standard-Model particles to better constrain
the $h_2$ mass or to reveal additional $h_2$ decay modes into long-lived
particles.
\end{abstract}


\section{Introduction\label{intro}}
The possibility of the Standard-Model (SM) Higgs field serving as the
portal to dark matter \cite{Patt:2006fw} has been extensively
phenomenologically studied in the past two decades. A viable scenario
involves a gauge singlet Higgs field which mixes with the SM Higgs field
through appropriate terms in the Higgs potential, resulting in a
dominantly SU(2)-doublet Higgs boson $h_1$ with mass 125 GeV and an
additional Higgs boson $h_2$ with a priori arbitrary mass
\cite{Schabinger:2005ei,Greljo:2013wja,Krnjaic:2015mbs}.  If the mixing angle is sufficiently
small, the couplings of the 125-GeV Higgs $h_1$ comply with their SM
values within the experimental error bars. The other Higgs boson $h_2$,
which is mostly gauge singlet, serves as a mediator to the Dark
Sector. In the simplest models the mediator couples to pairs of
dark-matter (DM) particles. In this paper we are interested in the
imprints of the described Higgs portal scenario on rare B meson decays
which can be studied in the new Belle II experiment. If the $h_2$ mass
is in the desired range below the $B$ mass, the decay of $h_2$ into a
pair of DM particles must necessarily be kinematically forbidden to
comply with the observed relic DM abundance
\cite{Greljo:2013wja,Krnjaic:2015mbs}. Phenomenological studies of the scenario were
recently performed in Refs.~\cite{Krnjaic:2015mbs,Winkler:2018qyg,Matsumoto:2018acr,Boiarska:2019jym,Filimonova:2019tuy}.

In this article we first revisit the calculation of the loop-induced
amplitude $b\to s h_2$. The literature on the topic employs a result
derived from the SM $\bar s b$-Higgs vertex with off-shell Higgs
\cite{Batell:2009jf}. However, it is known that this vertex is
gauge-dependent \cite{Botella:1986gf}.  This observation calls for a
novel calculation of the $\bar s b h_2$ vertex in an arbitrary $R_\xi$
gauge in order to investigate the correctness of the standard approach
and to understand how the gauge parameter $\xi$ cancels in physical
observables.  After briefly reviewing the model in Sec.~\ref{sec:m} we
present our calculation of the $\bar s b h_2$ vertex in
Sec.~\ref{sec:xi} and demonstrate the cancellation of the gauge
dependence for the two cases with on-shell $h_2$ and an off-shell $h_2$
coupling to a fermion pair, respectively. In Sec.~\ref{sec:p} we present
a phenomenological analysis with several novel aspects, such as a study
of the decay $B\to K^* h_2$ and a discussion of the lifetime information
inferred from data on $B\to K^{(*)} h_2[\to f \bar f]$ with a displaced
vertex of the $h_2$ decay into the fermion pair $f\bar f $. In
Sec.~\ref{sec:c} we conclude.

\section{Model\label{sec:m}}
A minimal extension of the SM with a real scalar singlet boson serving
as mediator to the Dark Sector involves the Higgs potential:
\begin{eqnarray}
  V &=& V_H+V_{H\phi}+V_\phi+\text{h.c.} \label{eq:v} \\
  \text{with}\qquad  V_H &=&  - \mu^{2} H^{\dagger} H +
                             \frac{\bar{\lambda}_0}{4}(H^{\dagger} H)^2,\nn\\
                 V_{H\phi}&=&\frac{\alpha}{2} \phi (H^{\dagger} H),\nn\\
      V_\phi &=&\frac{m^{2}}{2} \phi^{2}  + \frac{1}{4} \lambda_{\phi} \phi^{4},\nn
\end{eqnarray}
where $\phi$ denotes the scalar singlet field in the interaction basis, while
$H = \lt( G^{+}, (v+h + i G^{0} )/\sqrt{2} \rt)^T$ is the SM Higgs doublet.  We
minimize the scalar potential $V$ with respect to $\phi$ and $h$ and then choose
to express the mass parameters $\mu$ and $m$ in terms of corresponding vacuum
expectation values (vevs) $v_\phi$ and $v$, respectively:
\begin{eqnarray}
  \qquad  \mu_h^2&\equiv& \frac{\partial^2V}{\partial h^2}
                          \, =\,\frac{\bar{\lambda}_0 v^2}{2},\nn\\
  \mu_{h\phi}^2&\equiv& \frac{\partial^2V}{\partial h \partial\phi}
                        \,=\, \frac{\alpha v}{2},\nn\\
  \mu_\phi^2 &\equiv& \frac{\partial^2V}{\partial\phi^2}
                   \,=\, 2\lambda_\phi v_\phi^2-\frac{\alpha v^2}{4 v_\phi}\,.
\end{eqnarray}
The corresponding off-diagonal mass matrix is diagonalized with the introduction
of the mixing angle $\theta$
\begin{eqnarray}
\quad h = \cos \theta\,h_{1} - \sin \theta\, h_{2},\quad
\phi = \sin \theta\, h_{1} + \cos \theta\, h_{2}\,.
\end{eqnarray}
As mentioned in the introduction, we choose $h_2$ as the light mass eigenstate,
whose signatures we are primarily interested in, while $h_1$ corresponds to
the observed Higgs boson with mass $125\,\text{GeV}$.

An important Feynman rule for the calculation of the scalar penguin in
$R_\xi$ gauge is the one for the $G^+ G^- h_2$ vertex.  After diagonalization the
mass matrix we find\footnote{We express the Feynman rules using the conventions of
  the SM file in the \emph{FeynArts}\ \cite{Hahn:2000kx} package.}
\begin{eqnarray}
  \qquad  G^+ G^- h_1:\enskip\quad
              -i\frac{e m_{h_{1}}^2 \cos\theta}{2m_W \sin\theta_W},  \nn\\
\qquad  G^+ G^- h_2:\enskip\quad  \phantom{-} i\frac{e m_{h_{2}}^2 \sin\theta}{2m_W \sin\theta_W} .
  \label{eq:ggh2}  
\end{eqnarray}
One easily verifies that the rest of the vertices that are required for the studies
of low energy phenomenology are simple rescalings of the corresponding SM Higgs
vertices by the factor $(-\sin\theta)$. Note that the $G^+ G^- h_2$ vertex is not
found in the same way from the corresponding SM vertex, but in addition involves
the proper replacement of the SM Higgs mass by $m_{h_2}$. 

One could have included more terms in the scalar potential in \eq{eq:v} such as
$\phi^2 H^{\dagger} H$, however, such terms would not change the low-energy
phenomenology related to the process of our interest but would merely influence
the scalar self-interactions that we are currently not concerned with.
\boldmath 
\section{The $\bar{s} bh_2$ vertex in the $R_\xi$
  gauge\label{sec:xi}}
\begin{figure*}[tb]
	\begin{center}
		\subfigure[t][]{\includegraphics[width=0.25\textwidth]{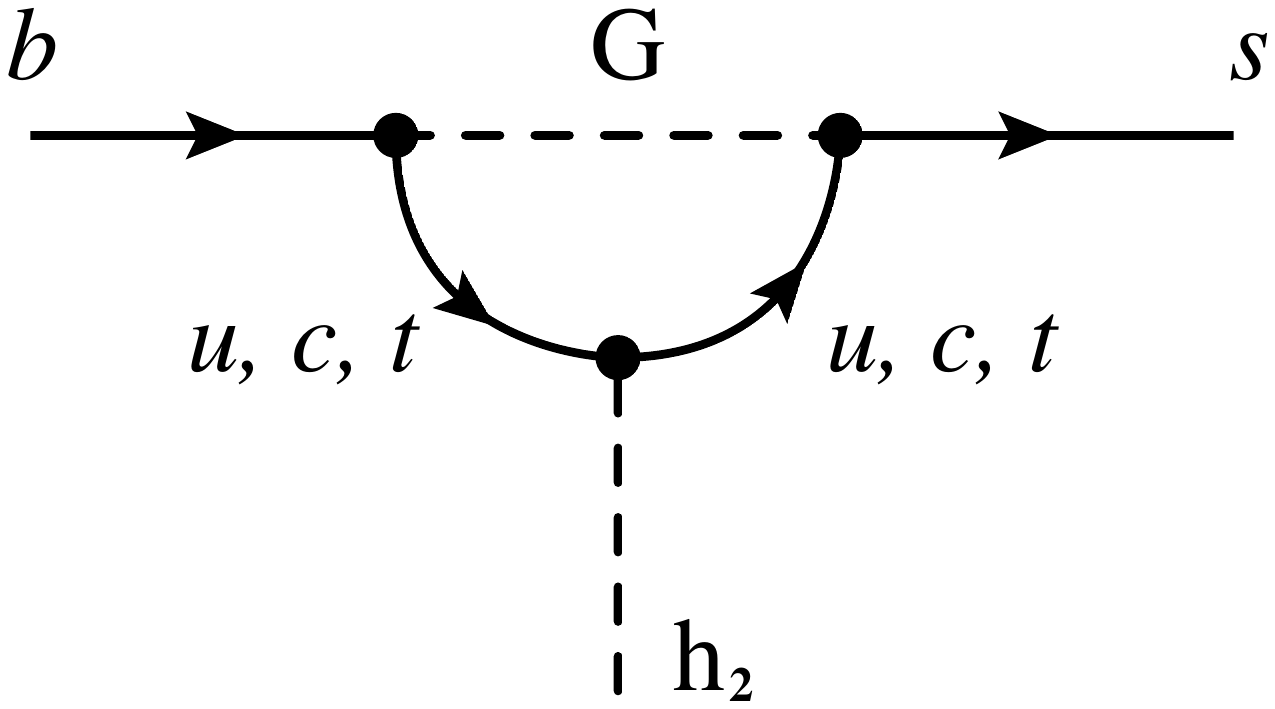}}
		\hspace{.6cm}
		\subfigure[t][]{\includegraphics[width=0.25\textwidth]{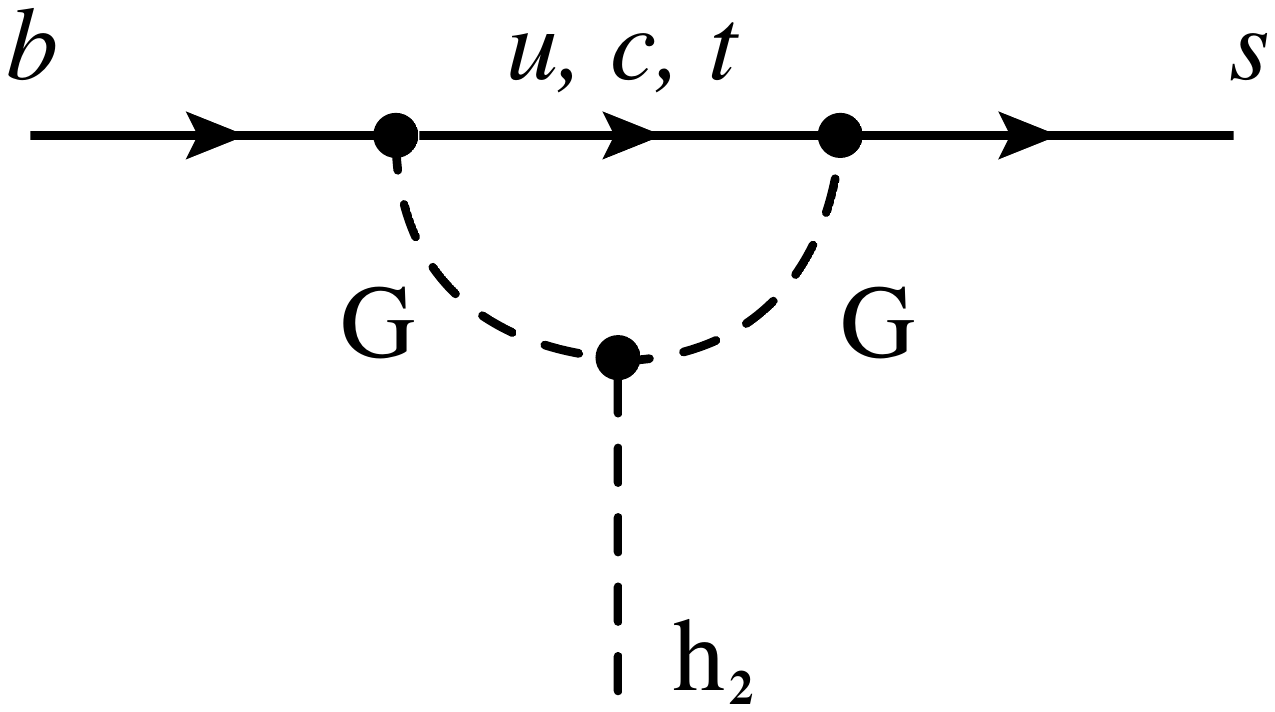}}
		\hspace{.6cm}
		\subfigure[t][]{\includegraphics[width=0.25\textwidth]{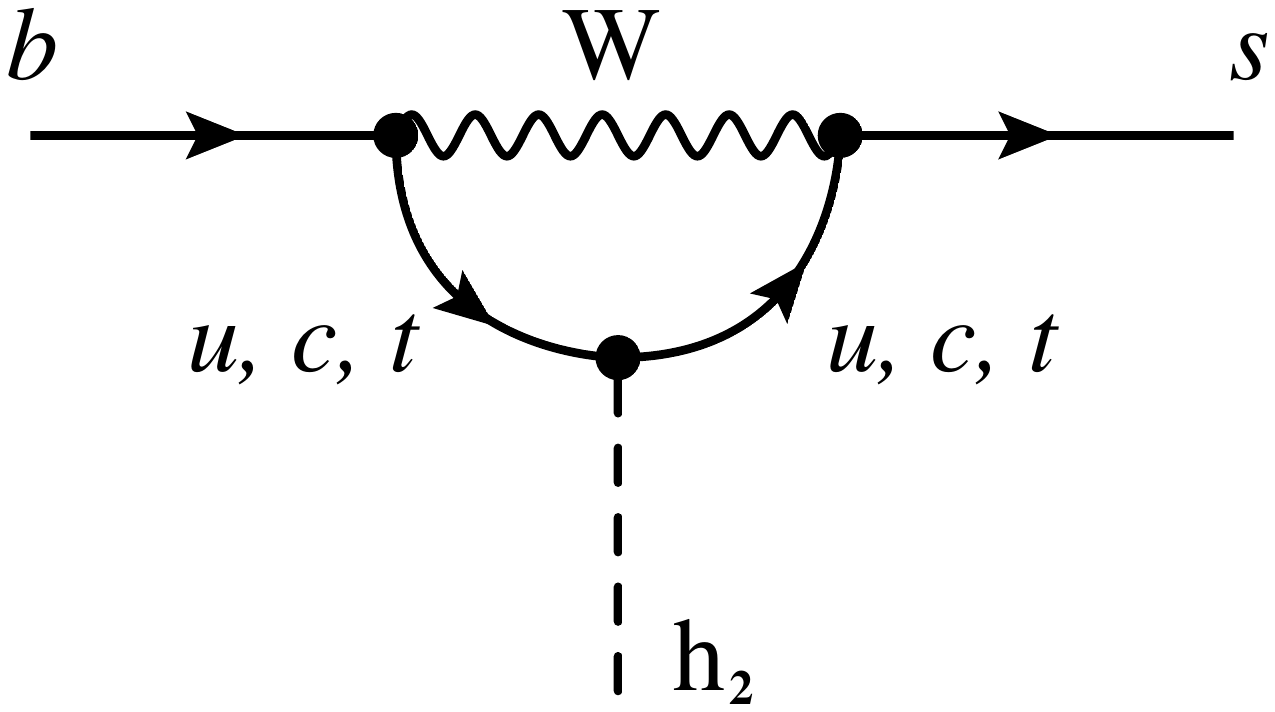}}\\
		\hspace{.6cm}
		\subfigure[t][]{\includegraphics[width=0.25\textwidth]{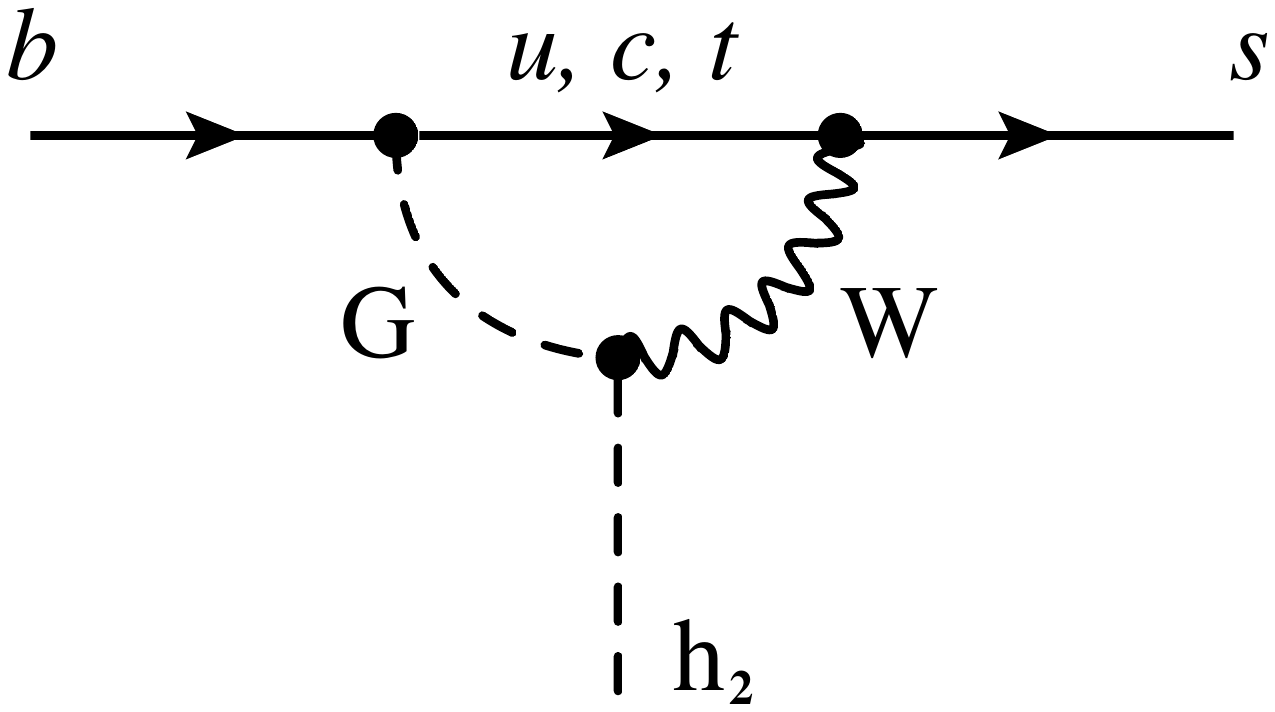}}
		\hspace{.6cm}
		\subfigure[t][]{\includegraphics[width=0.25\textwidth]{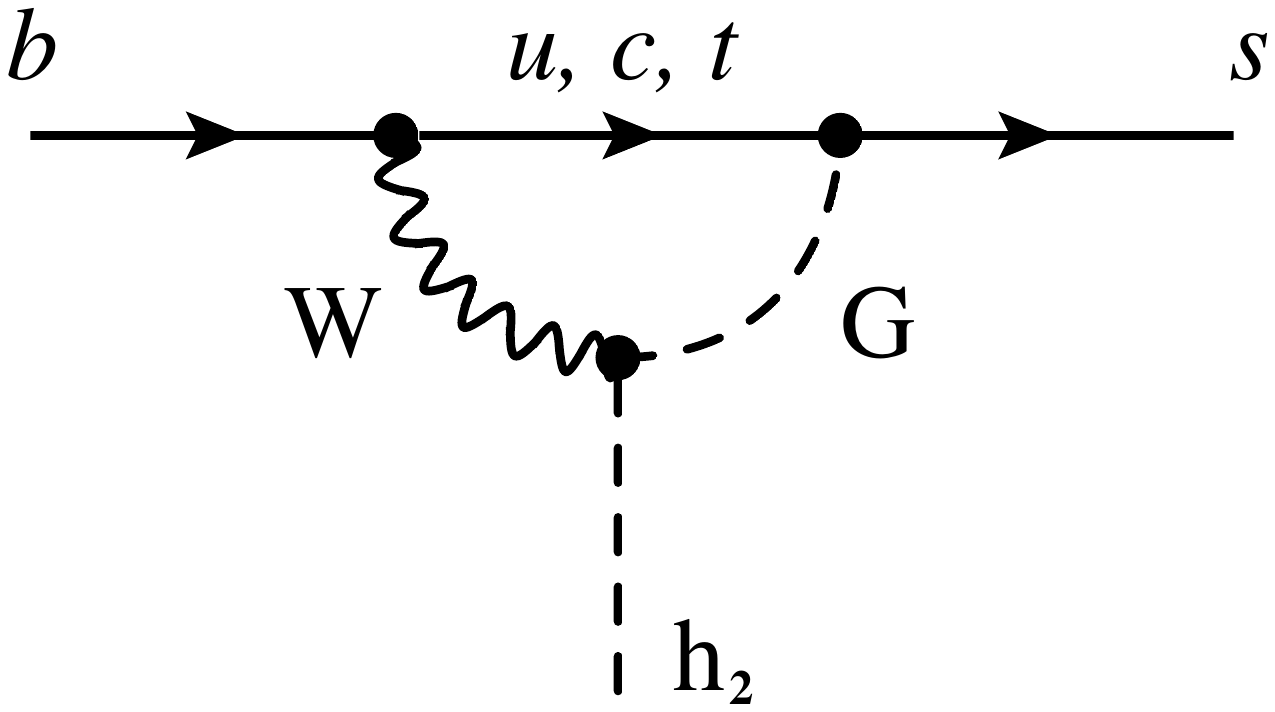}}
		\hspace{.6cm}
		\subfigure[t][]{\includegraphics[width=0.25\textwidth]{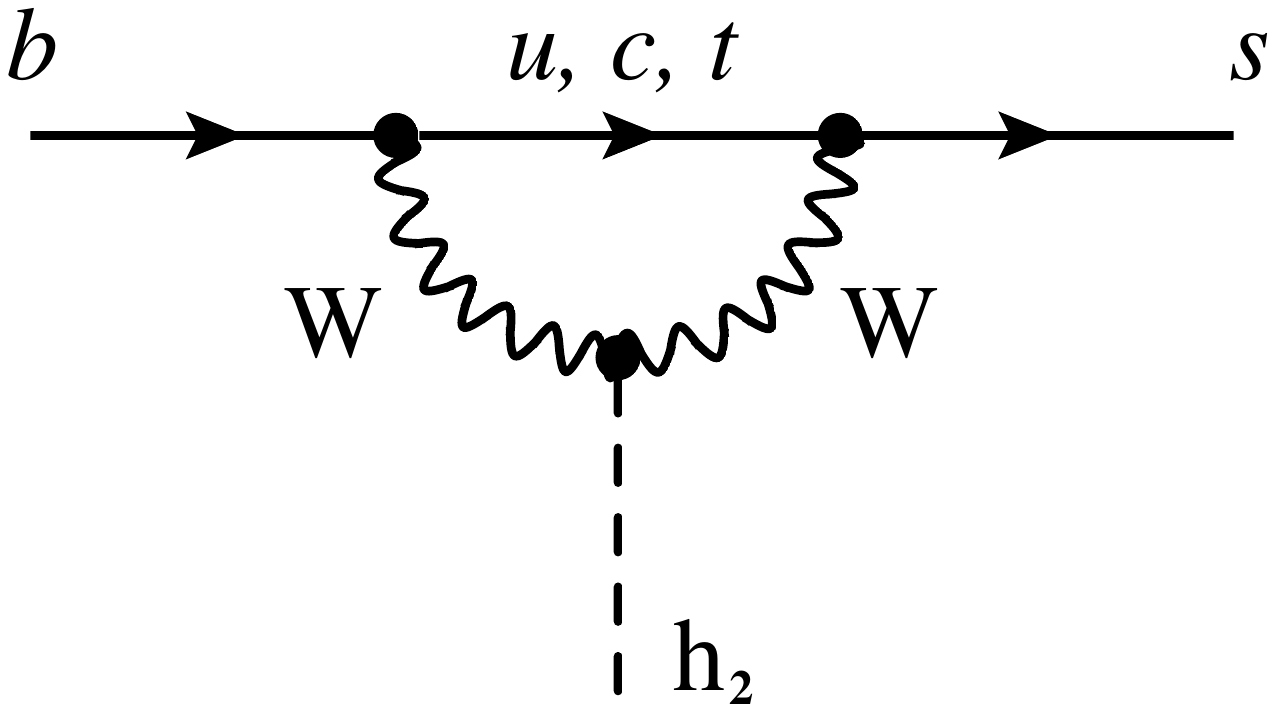}}\\
		\hspace{.6cm}
		\subfigure[t][]{\includegraphics[width=0.25\textwidth]{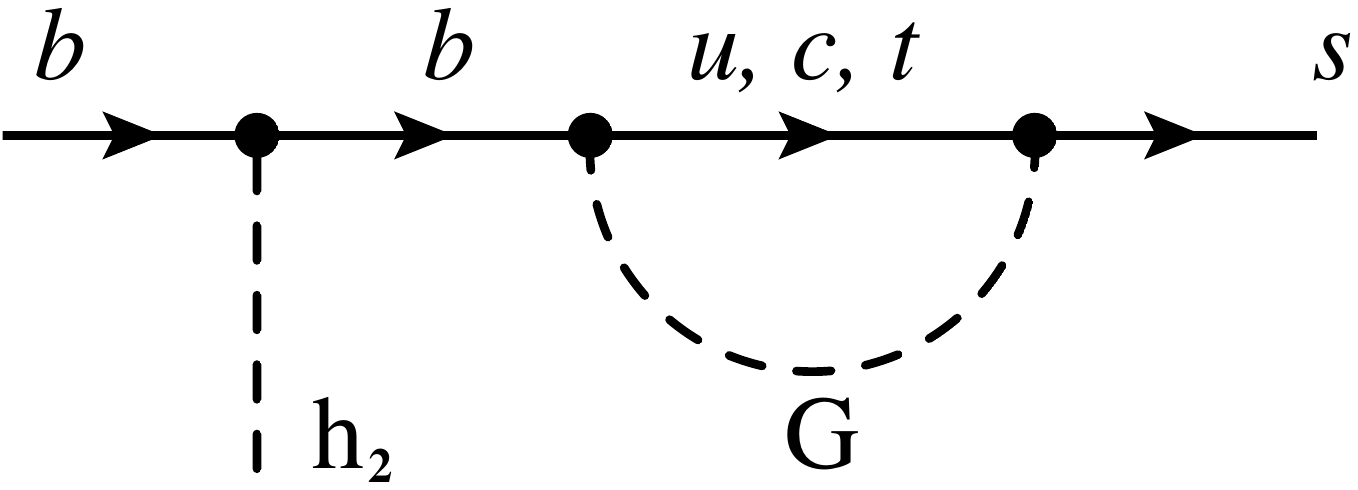}}
		\hspace{.6cm}
		\subfigure[t][]{\includegraphics[width=0.25\textwidth]{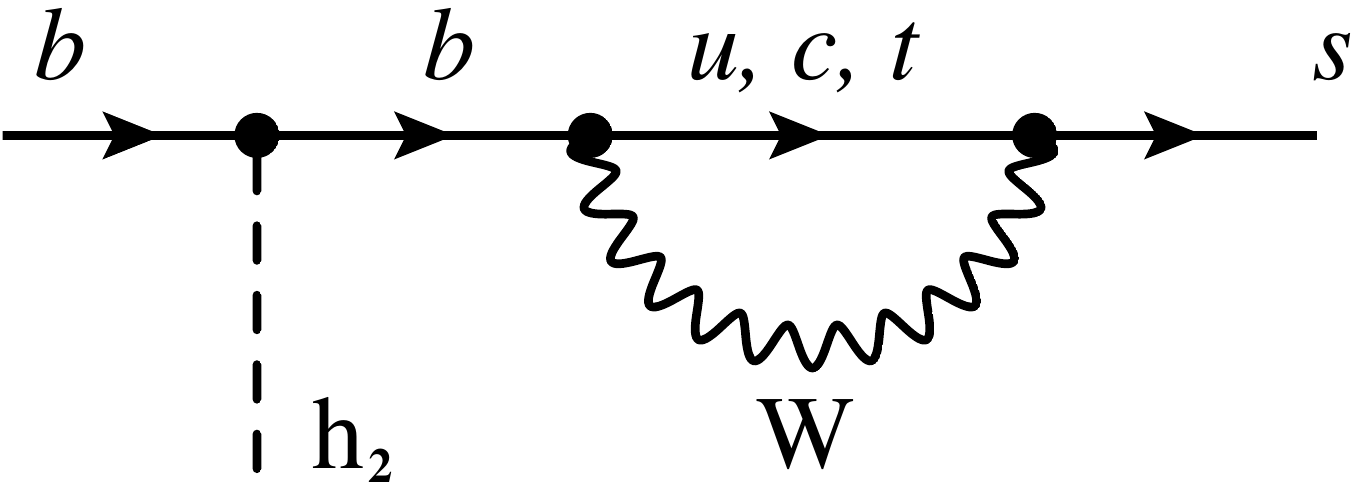}}
         \end{center}
         \caption{One-loop
           diagrams contributing to $b\to s h_2$ 
           in $R_\xi$ gauge.}
	\label{Fig:diags}
~\\[-3mm]\hrule
\end{figure*}

\unboldmath%
We employ a general $R_\xi$ gauge for the calculation of the Feynman
diagrams contributing to the $\bar{s}\text{-}b\text{-}h_2$ vertex. We
further use the \emph{FeynArts}\ package \cite{Hahn:2000kx} for
generating the amplitudes and the \emph{FeynCalc}\
\cite{Mertig:1990an,Shtabovenko:2016sxi,Shtabovenko:2020gxv},
\emph{Package-X}\ \cite{Patel:2015tea}, and \emph{FeynHelpers}\
\cite{Shtabovenko:2016whf} packages to evaluate the analytic expressions
for the Feynman diagrams.  Neglecting the mass of the external $s$
quark, we encounter the diagrams shown in \fig{Fig:diags}. In our final
result we will also neglect the masses of the internal up and
charm quarks.  While the expressions for individual diagrams contain
ultraviolet poles, the final result is UV convergent due to the
Glashow-Iliopoulos-Maiani mechanism.

In order to elucidate the gauge independence of the physical quantities,
we set the $h_2$ boson off the mass shell. In a first step we
present the results in terms of the scalar loop functions $B_0,C_0$ of the
Passarino-Veltman (PV) basis, keeping exact dependences on all momenta
and masses.  For the final goal to calculate the low-energy Wilson
coefficient governing the decay process $b\to s\ h_2$ this appears
unnecessary, but it turns out that the expression in terms
of the PV basis is compact and most suitable for studying
the gauge-independence of the physical quantities.

We decompose each diagram $\mathcal{A}_i$ as
$\mathcal{A}_i= \tilde{\mathcal{A}}_{i} + \mathcal{A}_{i}^{(\xi)}$, with
the second term $\mathcal{A}_{i}^{(\xi)}$ comprising all terms which
depend on the $W$ gauge parameter $\xi$.  The expressions for
$\tilde{\mathcal{A}}_i$ are collected in \ref{Results}.
The results for the gauge-dependent pieces of the individual diagrams are
rather lengthy, so we only provide the total sum
\begin{eqnarray}
  \;\sum_i \mathcal{A}_i^{(\xi)} & = & \sin\theta
            \dfrac{ m_b m_t^2 }{ 8\pi^2 v^3 (m_b^2-p_{h_2}^2) }
                                               (p_{h_2}^2-m_{h_2}^2)
                                               \cdot \nn\\
  &&\!\!\!\bigg[B_0(p_{h_2}^2, m_W^2 \xi, m_W^2 \xi)-B_0(m_b^2, m_t^2, m_W^2
     \xi) \nn \\
 &&\quad +\,(p_{h_2}^2-m_b^2+m_t^2-m_W^2\xi)\cdot \nn \\
  &&\qquad C_0(0,m_b^2,p_{h_2}^2, m_W^2\xi\,, m_t^2,m_W^2\xi )\bigg]\,,
  \label{Gauge-dependent}
\end{eqnarray}
with $\lambda_t = V_{tb}V^\ast_{ts}$. Here and in the following we
suppress the Dirac spinors for the $b$ and $s$ quarks. 
It follows from the expression
above that the gauge-dependent contribution $\mathcal{A}^{(\xi)}$
vanishes for the case of an on-shell scalar boson, which confirms the gauge
independence of the corresponding physical on-shell amplitude. We write
the total $\bar{s} bh_2$ vertex
$\mathcal{A} = \sum_i (\tilde{\mathcal{A}}_i + \mathcal{A}_i^{(\xi)}) $ (with on-shell quarks and
off-shell $h_2$) as
\begin{eqnarray}
 \quad && \mathcal{A} = G(p_{h_2}^2, m_{h_2}^2) +
     (p_{h_{2}}^{2} - m_{h_{2}}^{2}) F(\xi, p_{h_2}^2) , 
\end{eqnarray}
with the second term equal to the expression in \eq{Gauge-dependent}.
We note that $ F(\xi, p_{h_2}^2) $ does not depend on $m_{h_2}$.  While
the cancellation of $\xi$ from $\mathcal{A} $ is obvious for an on-shell
$h_2$, i.e.\ for the decay $b\to s\ h_2$, this feature is not
immediately transparent for the case in which an off-shell $h_2$ decays
into a pair of other particles. In such scenarios the gauge dependence
is cancelled by other diagrams. Here we exemplify the cancellation of
the gauge parameter for a model in which our mediator $h_2$ couples to
a pair of invisible final state fermions:
\begin{equation}
\qquad \mathcal{L}_{\phi \chi \chi} \;=\; \lambda_\chi \phi \overline{\chi} \chi\,,
\end{equation}
meaning that $h_2$ in $b\to s\ h_2 [\to \overline{\chi} \chi]$ is
necessarily off-shell \cite{Krnjaic:2015mbs}.  In order to find the
cancellation of the gauge parameter we must also consider the
diagrams corresponding to  $b\to s\ h_1 [\to \overline{\chi} \chi]$
involving the heavy SM-like state $h_{1}$.
The amplitudes involving the $h_2$ and $h_1$ propagators are
proportional to $- \sin \theta$ and to $\cos \theta$, respectively:
\begin{equation}
\begin{split}
\qquad \mathcal{A}_{b\text{-}s\text{-}h2}  \sim  - \sin \theta,\qquad\qquad
\mathcal{A}_{b\text{-}s\text{-}h1} \sim  \cos \theta,
\end{split}
\end{equation}
while the vertices $\mathcal{V}_{h_{1,2} \chi \chi } $
involving the coupling of the dark-matter fermion to
the scalar bosons depend on $\theta$ as
$\mathcal{V}_{h_{1} \chi \chi } \sim \sin \theta$ and
$\mathcal{V}_{h_{2} \chi \chi } \sim \cos \theta$. The
$b\to s\ h_{1,2} [\to \overline{\chi} \chi]$ amplitudes $\mathcal{A}_{h_{1,2}}$ can be
schematically written as
\begin{eqnarray}
  \quad\mathcal{A}_{h_{2}} &=&  - \lambda_\chi
                                \sin \theta \cos \theta \left(F(\xi, p^2) +
                    \frac{G(p^2, m_{h_{2}}^{2})}{p^{2} - m_{h_{2}}^{2}}\right), \\
  \quad\mathcal{A}_{h_{1}} &=&  \phantom{-}\lambda_\chi
                                \sin \theta \cos \theta \left( F(\xi, p^2) +
        \frac{G(p^2, m_{h_{1}}^{2})}{p^{2} - m_{h_{1}}^{2}}\right),
\end{eqnarray} 
where $p^2$ denotes the square of the momentum transferred to the
fermion pair.  By adding the two amplitudes one verifies the cancellation of the
gauge-depend\-ent part $ F(\xi, p^2)$. If one considers processes with
off-shell $h_{1,2}$ exchange to SM fermions, such as in $b\to s\tau^+
\tau^-$ with e.g.\ $m_{h_2}>m_b$, also box diagrams are needed for the
proper gauge cancellation as found in Ref.~\cite{Botella:1986gf} for the
SM case. 

We now proceed to integrate out the top quark and W boson within the
gauge independent contribution
$\tilde{\mathcal{A}}\equiv \sum_i \tilde{\mathcal{A}}_i$ to obtain the
Wilson coefficient:
\begin{eqnarray}
\mathcal{L}_{\text{eff}}&=&C_{h_2 s b}\, h_2\, \overline{s} P_R
b+\text{h.c.}, \label{Leff}\\
  \qquad\qquad
  C_{h_2 s b} &=& -\frac{3\,\sin\theta\,\lambda_t \,m_b\,m_t^2}{16\,\pi^2\,v^3}\,,
\end{eqnarray}
where $v\simeq 246\,\text{GeV}$ is the vacuum expectation value of the
Higgs doublet.  This result agrees with Ref.~\cite{Winkler:2018qyg}, whereas it agrees with Refs.~\cite{Krnjaic:2015mbs} and \cite{Batell:2009jf} up to the sign.\footnote{The result in Ref.~\cite{Krnjaic:2015mbs} has the sign opposite to us, while we cannot conclude which sign convention is used in Ref.~\cite{Batell:2009jf}.}

The procedure to multiply the SM result for the $\bar s b$-Higgs vertex
by $-\sin\theta $ to find the $\bar{s} bh_2$ vertex is not correct in an
$R_\xi$ gauge (nor for the special cases $\xi=0$ or $\xi=1$ of the Landau and 't
Hooft-Feynman gauges) because of the subtlety with the $G^\pm$ vertices
in \eq{eq:ggh2}. However, the missing terms are suppressed by higher
powers of $m_{h_2}^2/M_W^2$ and do not contribute to the effective
dimension-4 lagrangian in \eq{Leff}.

\begin{figure*}[tb]
	\begin{center}
		\subfigure{\includegraphics[width=0.55\textwidth]{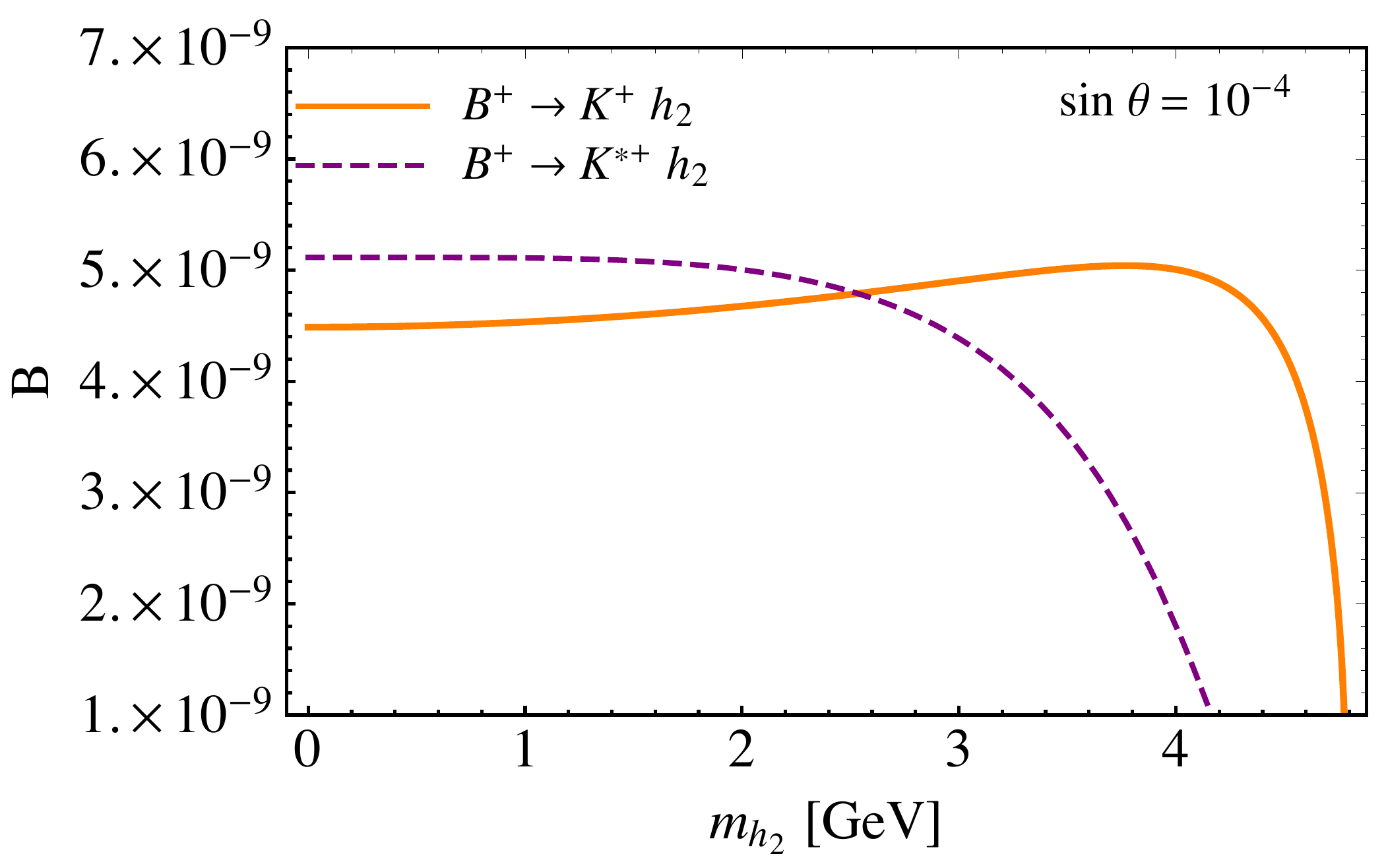}}
         \end{center}
         \caption{Comparison of the branching fractions of
           $B^+\to K^+ h_{2}$ (thick orange curve) and $B^+\to K^{\ast +} h_{2}$ (dashed purple curve) for $\sin\theta=10^{-4}$.}
	\label{Fig:BranFrac}
~\\[-3mm]\hrule
\end{figure*}
\section{Phenomenology\label{sec:p}}
The experimental signature $B\to K\, h_{2}$ permits the
determination of $m_{h_2}$ from the decay kinematics, while the other
relevant parameter of the model, $\sin\theta$, can be determined from
the measured branching ratio $B(B\to K\, h_{2})$. With increasing
$m_{h_2}$ more $h_2$ decay channels open and the $h_2$ lifetime may be
in a favourable range allowing the $h_2$ to decay within the Belle II
detector. This scenario has a characteristic displaced-vertex signature
which is highly beneficial for the experimental analysis. Higgs-portal
signatures at $B$ factories have been widely studied
\cite{Krnjaic:2015mbs,Winkler:2018qyg,Filimonova:2019tuy,Kamenik:2011vy,Schmidt-Hoberg:2013hba,Clarke:2013aya,Sierra:2015fma}.
In this paper we briefly revisit
the recent analyses of Refs.~\cite{Winkler:2018qyg,Filimonova:2019tuy}
and complement them with novel elements: Firstly, we pre\-sent a novel analysis of the decay mode $B\to K^{*}(892) h_2$ in comparison to $B\to K h_2$.  Secondly, we highlight the benefits of the lifetime information
which can be obtained from the displaced-vertex data. Thirdly, we
present a new result of the number of $B\to K h_2[\to f]$ events
(with $f$ representing a pair of light particles) expected at Belle II as
a function of the relevant $B\to K h_2$ and $h_2\to f$ branching ratios.

In our study of $B\to K h_2$ and $B\to K^\ast h_2$ with subsequent
decay of $h_2$ into a visible final states with displaced vertex we
restrict ourselves to the case $m_{h_2}>2m_{\mu}$. 
While the leptonic decay rate is given by the simple formula
\begin{equation}
  \quad
  \Gamma(h_2\to \ell\ell) =
  \sin^2\theta \frac{G_F m_{h_{2}} m_{\ell}^2}{4\sqrt{2}\pi}
  \bigg(1-\frac{4m_\ell^2}{m_{h_{2}}^2}\bigg)^{3/2}\,,
\label{eq:s}
\end{equation}
the calculation of the decay rate into an exclusive hadronic final state
is challenging.  Different calculations of $ \Gamma(h_2\to \pi\pi)$ and
$ \Gamma(h_2\to KK)$
\cite{Voloshin:1985tc,Truong:1989my,Donoghue:1990xh,Monin:2018lee} employing chiral
perturbation theory have been clarified, updated and refined in
Ref.~\cite{Winkler:2018qyg} and we use the results of this reference.
In the region with $m_{h_{2}}> 2\,\text{GeV}$ the inclusive hadronic
decay rate can be reliably calculated in perturbation theory
\cite{Grinstein:1988yu}.

Analyses with fully visible final states $K^{(\ast)}f$ can also be done at LHCb~\cite{Aaij:2016qsm,Aaij:2015tna}.

\subsection{$B \to K h_2$}
The branching ratio of $B\to K h_2$  is
\begin{eqnarray}
\;  B(B\to K h_2) = \frac{\tau_B}{32\pi m_B^2}\vert C_{h_2 s
    b}\vert^2\bigg(\frac{m_B^2-m_K^2}{m_b-m_s}\bigg)^2\cdot \nn\\
  f_0(m_{h_{2}}^2)^2
  \frac{\lambda(m_B^2, m_K^2,m_{h_{2}}^2)^{1/2}}{2 m_B}\,,\label{BrtoK}
\end{eqnarray}
where $\lambda(a,b,c) = a^2+b^2+c^2 - 2(ab+ac+bc)$, and the scalar form
factor $f_0(q^2)$ is related to the desired scalar hadronic matrix element as
\begin{equation}
  \qquad \langle K\vert\bar{s}b\vert B\rangle =
  \frac{m_B^2-m_K^2}{m_b-m_s}\,f_0(q^2)\,,
\end{equation}
where $q=p_B-p_K$. For this form factor we use the QCD lattice
result of  Ref.~\cite{Bailey:2015dka} (see also
\cite{Bouchard:2013pna}).

The reach of the Belle II experiment for the process $B\to K h_2$ was
recently studied in Ref.\cite{Filimonova:2019tuy}. This investigation
involves a study of the detector geometry and we present a novel study
in \ref{App:Events}. For the evaluation of the number of events we use
the formula~\eqref{Events2}. Our evaluation of the sensitivities
corresponds to $5\cdot 10^{10}$ produced $B\bar{B}$ meson pairs,
where $B$ represents both $B^+$ and $B^0$, at
$50\,\text{ab}^{-1}$ of data at Belle II experiment \cite{Kou:2018nap}.

\begin{figure*}[tb]
\hrule\medskip
	\begin{center}
		\subfigure{\includegraphics[width=0.55\textwidth]{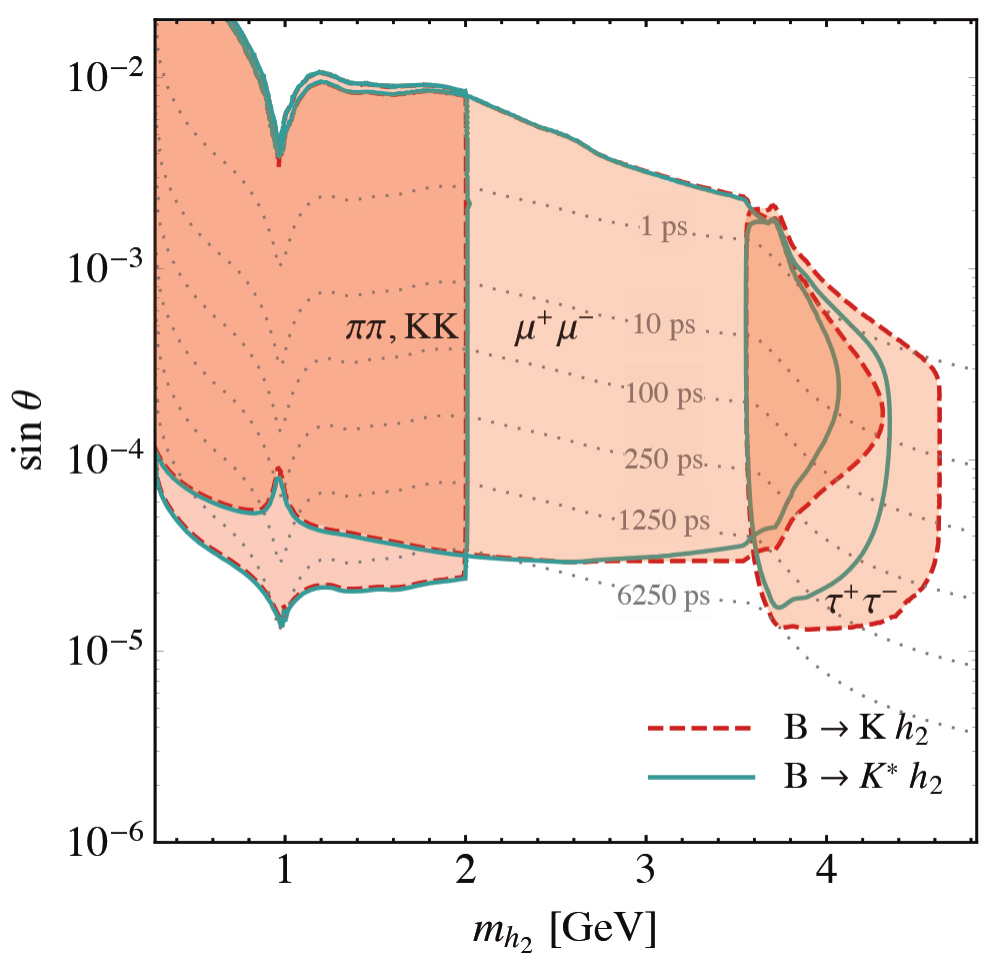}}
         \end{center}
         \caption{Parameter regions that correcpond to three or more
           events of $B\to K h_{2}\,(\to f)$,
           $f = (\pi\pi + K K), \mu^+\mu^-, \tau^+\tau^-$ are shaded in
           red and bounded by the dashed red contours. Analogous regions
           for $B\to K^\ast h_2$ are presented by the dark green
           contour. We summed over the number of events in the
           decays of $B^{+}$, $B^-$, $B^0$, and $\bar{B}^0$. The dotted
         lines are contours of constant $h_2$ proper lifetime.}
	\label{Fig:PlotA}
~\\[-3mm]\hrule
\end{figure*}

The parameter regions that correspond to three or more displaced vertex
events of any of the final state signatures in $B\to K (h_{2}\to f)$,
$f = (\pi\pi + K K), \mu\mu, \tau\tau$ within the Belle II detector are
displayed by the dashed red contours in figure \ref{Fig:PlotA}. The
number of events involve the summation over the decays of $B^+$,
$B^0$ and the corresponding charge-conjugate mesons. Following
Ref.~\cite{Filimonova:2019tuy}, we display the regions in which the
$\pi\pi, KK$ final states occur as well as the region above the $\tau$
lepton threshold within the same plot.  We show the contours of the
proper lifetime of the scalar mediator within the same parameter space
and encourage our experimental colleagues to include the lifetime
information in the following ways: In a first step one may assume the
minimal model adopted in this paper and use the lifetime measurements as
additional information on $m_{h_2}$ and $\sin \theta$.
E.g.\ if $h_2$ is light enough so that the only relevant decay channel is
$h_2\to \mu^+\mu^-$, the lifetime is the inverse of the width in
\eq{eq:s}. Thanks to the strong dependence on $m_{h_2}$ the lifetime
information will improve the determination of $m_{h_2}$ inferred from
the $B\to K h_2$ decay kinematics once $\sin \theta$ is fixed from
branching ratios. With more statistics one can go a step further and use
the lifetime information to verify or falsify the model. Even if all
$h_2$ couplings to SM particles originate from the SM Higgs field through
mixing, a richer singlet scalar sector can change the $h_2$
lifetime. Consider an extra gauge singlet scalar field $\tilde\phi$
coupling to $\phi$ in the potential in \eq{eq:v} giving rise to a third
physical Higgs state $h_3$. If $h_3$ is sufficiently light,
$h_2\to h_3h_3$ is possible. Through $\tilde\phi$--$H$ mixing the new
particle $h_3$ will decay back into SM particles, but
the lifetime can be so large that $h_2\to h_3h_3$ is just
a missing-energy signature. Then the only detectable effect of the extra
$h_2\to h_3h_3$ mode is a shorter $h_2$ lifetime. If
measured precisely enough, the lifetime will permit to determine the
decay rate of $h_2\to h_3h_3$ and thereby the associated
coupling constant. Alternatively, one may fathom a model in which
$h_2$ decays into a pair of sterile neutrinos which decay back to SM fermions.

\subsection{$B \to K^{\ast} h_2$}
We include in our analysis the decay of $B$ meson that involves the final
state vector meson $K^\ast$ and has the branching fraction
\begin{eqnarray}
  \quad
  B(B\to K^\ast h_2) &=& \frac{\tau_B}{32\pi m_B^2}\vert C_{h_2 s b}\vert^2
\frac{A_0(m_{h_2}^2)^2}{(m_b+m_s)^2} \cdot \nn\\
&&\quad \frac{\lambda(m_B^2, m_{K^\ast}^2, m_{h_2}^2)^{3/2}}{2 m_B}\,.\label{BrtoKst}
\end{eqnarray}
The form factor $A_0(q^2)$ is related to the desired pseudoscalar
hadronic matrix element as
\begin{equation}
\langle K^\ast (k, \epsilon)\vert \bar{s}\gamma_5 b\vert B(p_B)\rangle = \frac{2\,m_{K^\ast}\,\epsilon^\ast\cdot q}{m_b+m_s} A_0(q^2)\,, 
\end{equation}
where $\epsilon$ is a polarization vector of $K^\ast$ and $q=p_B-k$. For this form factor we use the combination of results from lattice QCD \cite{Horgan:2013hoa} and QCD sum rules \cite{Straub:2015ica} as provided in Ref.~\cite{Straub:2015ica}.

$ B(B\to K^\ast h_2)$ is comparable in size to $ B(B\to K h_2)$ for
masses up to $\sim 2\,\text{GeV}$ (see \fig{Fig:BranFrac}), and is
suppressed as the mass $m_{h_{2}}$ approaches the kinematic
endpoint. This is the result of the additional power of the kinematic
function $\lambda$ in \eq{BrtoKst} that comes from the
contribution of the longitudinal $K^\ast$ polarization. It follows from
angular momentum conservation that this is the only contributing
polarization. The combination of the experimental data from both
processes will be required in order to discriminate the spin-0 vs.\
spin-1 hypotheses
in case of a discovery. E.g.\ the mediator with spin 1 involves
a different dependence of the rate on the mediator's mass
and comes with a dramatic suppression of the decay rate with $K$ 
in the final state if the mediator is light. The decay  $B \to K^\ast h_2$ has been studied before in Ref. \cite{Boiarska:2019jym}, in which a plot similar to our Fig. \ref{Fig:BranFrac} is presented
for the sum of several vector resonances.  Our analysis of Belle II opportunities is new compared to Ref. \cite{Boiarska:2019jym} which focuses on LHC, ShiP, and DUNE. Refs. \cite{Boiarska:2019jym,Filimonova:2019tuy} further study the fully inclusive decay $B \to X_s h_2$.

The kinematic suppression close to the endpoint implies that the number
of $B\to K^\ast h_2(\tau\tau)$ events will be much smaller relative to
the case of the final state with $K$. We display the corresponding parameter
region corresponding to $K^\ast$ events with the dark green contour
in \fig{Fig:PlotA}.

 \begin{figure*}[tb]
	\begin{center}
		\subfigure{\includegraphics[width=0.55\textwidth]{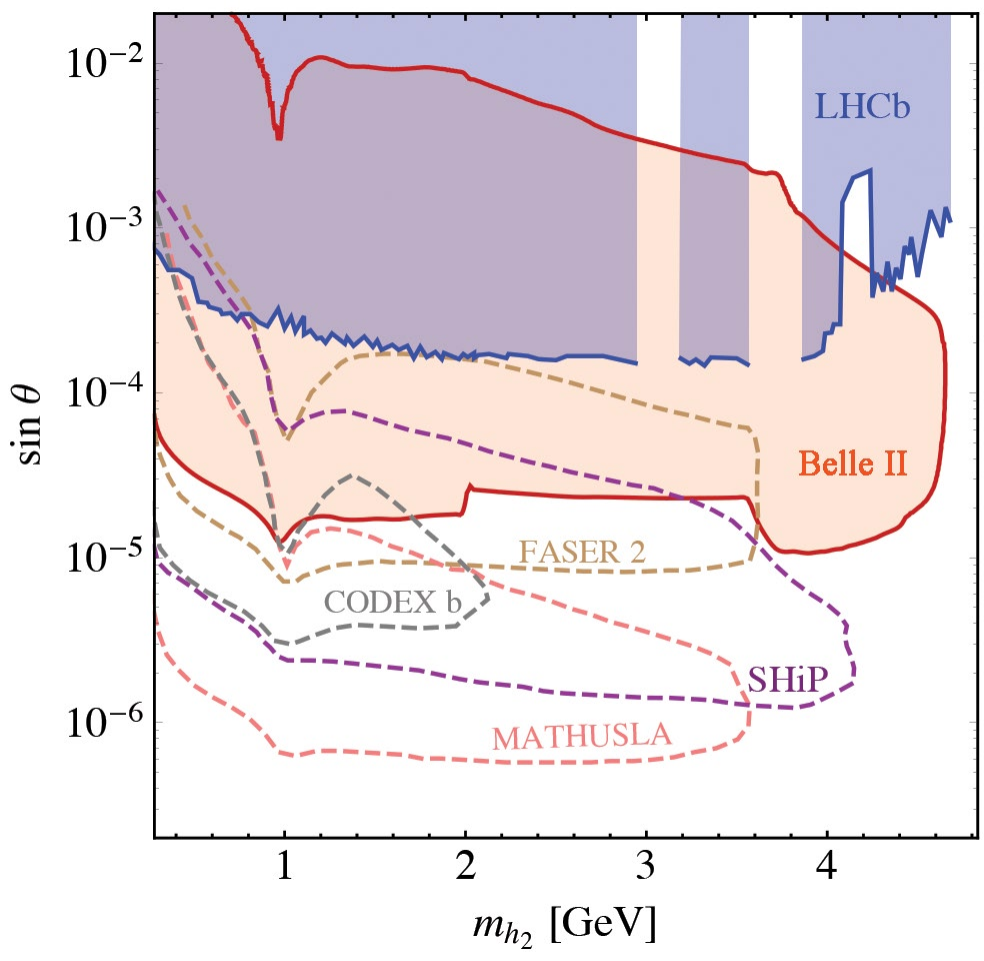}}
         \end{center}
         \caption{Combined sensitivity of the Belle II experiment to
           displaced vertices of $h_2$ including both $B\to K h_{2}$ and
           $B\to K^\ast h_{2}$ and decays of $h_2$ to
           $(\pi\pi + K K), \mu^+\mu^-, \tau^+\tau^-$ are shown with the
           filled red region, and compared to the search limit of LHCb
           \cite{Aaij:2016qsm} (shaded blue) and projected sensitivities
           by other proposed experiments, Mathusla~\cite{Evans:2017lvd}
           (pink), SHiP~\cite{Alekhin:2015byh}, CODEX
           b~\cite{Gligorov:2017nwh} (gray) and FASER
           2~\cite{Ariga:2018uku} (brown).}
	\label{Fig:combination}
~\\[-3mm]\hrule
\end{figure*}
In \fig{Fig:combination} we compare the reach of the Belle II experiment
to displaced vertices of $h_2$ including both $B\to K h_{2}$ and
$B\to K^\ast h_{2}$ processes and decays of $h_2$ to
$(\pi\pi + K K), \mu^+\mu^-, \tau^+\tau^-$ with the existing search
limit of the LHCb experiment~\cite{Aaij:2016qsm}.\footnote{We use 
  the result of Ref.~\cite{Winkler:2018qyg} for the LHCb search limit on
  $B(B\to K h_2[\to\mu^+\mu^-])$.} We also compare to projected
sensitivities of other proposed experiments,
Mathusla~\cite{Evans:2017lvd}, SHiP~\cite{Alekhin:2015byh}, CODEX
b~\cite{Gligorov:2017nwh} and FASER 2~\cite{Ariga:2018uku}.

\section{Conclusions}\label{sec:c}
We have clarified the cancellation of gauge-dependent terms appearing in
the $\bar{s} bh_2$ vertex in the standard Higgs portal model with a
singlet mediator to the Dark Sector. We have further updated the
$b\to s h_2$ phenomenology to be studied at the Belle II detector, with
a novel consideration of $B\to K^* h_2$ complementing the previously
studied decay $B\to K h_2$.  Decays like
$B\to K^{(*)} h_2[\to \mu^+\mu^-]$ with a displaced vertex permit the
measurement of the $h_2$ lifetime. It is shown how this measurement will
further constrain the two relevant parameters $m_{h_2}$ and $\sin\theta$
of the model. Both the lifetime information and the combined study of
$B\to K^* h_2$ and $B\to K h_2$ permit the discrimination of the studied
Higgs portal from other Dark-Sector models. Another result
of this paper is a new calculation of the expected number of
$B\to K^{*} h_2[\to f]$ events as a function of the $B\to K h_2$ and
$h_2\to f$ branching ratios for the Belle II detector.

\begin{acknowledgements}
  We are grateful for helpful discussions with Teppei Kitahara, Felix
  Metzner, Vladyslav Shtabovenko and Susanne
  Westhoff and thank the authors of Ref.~\cite{Filimonova:2019tuy}
   for confirming our result in \ref{App:Events}.
  We further thank Ulises J.\ Salda\~na-Salazar
    for participation in the early  stages of the project. This work is
  supported by BMBF under grant \emph{Verbundprojekt 05H2018 (ErUM-FSP
    T09) - BELLE II: Theoretische Studien zur Flavourphysik}. A.K.\
  acknowledges the support from the doctoral school \emph{KSETA}\ and
  the \emph{Graduate School Scholarship Programme}\ of the \emph{German
    Academic Exchange Service (DAAD)}.
  
\end{acknowledgements}
\appendix
\section{Results of the loop calculation}\label{Results}
In this appendix we present the results for the $\xi$-in\-de\-pen\-dent pieces
$\tilde{\mathcal{A}}_{i}$ corresponding to the individual Feynman diagrams
shown in \fig{Fig:diags}:

\begin{align}
 \nonumber \tilde{\mathcal{A}}_{(a)} = - \sin\theta\,\dfrac{\lambda_t\, m_b m_t^2}{8\pi^2 v^3}\frac{p_{h_{2}}^{2} - 2 m_{t}^{2}}{m_{b}^{2} - p_{h_{2}}^{2}} \, B_{0}(p_{h_{2}}^{2}, m_{t}^{2} , m_{t}^{2})\,
 \end{align}
\begin{align}
 \tilde{\mathcal{A}}_{(b)} &= 0\,,
\end{align}
\begin{align}
\nonumber \tilde{\mathcal{A}}_{(c)} = & -\sin\theta\dfrac{\lambda_t\,m_t^2}{16\,\pi^2\,m_b v^3}\dfrac{1}{m_b^2-p_{h_{2}}^{2}}\cdot\\
\nonumber &\bigg\{\big[-m_b^2\big(m_W^2 (4D + 5 x - 9) + p_{h_{2}}^{2}\big) + 3m_b^4 \\
\nonumber &+ m_W^2 p_{h_{2}}^{2}(x-1) \big] B_{0}(m_b^2, m_t^2, m_W^2) \\ 
\nonumber &+ 2m_b^2 m_W^2\big(m_b^2(2-x)+2m_W^2 (x-1) (2+x)\\
\nonumber & - p_{h_{2}}^{2}\big)C_{0}(0, m_b^2, p_{h_{2}}^{2}, m_t^2, m_W^2, m_t^2)\\ \nonumber
&-4(D-2) m_b^2 m_W^2 B_{0}(p_{h_{2}}^2, m_t^2, m_t^2) \\
 &+ \dfrac{2m_W^2 (m_b^2-p_{h_{2}}^{2})}{D-2} B_{0}(0,m_W^2, m_W^2) \bigg\}\,,
\end{align}
\begin{align}
\nonumber \tilde{\mathcal{A}}_{(d)} =& -\sin\theta\dfrac{\lambda_t\,m_b}{8\pi^2 v^3}\big((m_t^2-2m_W^2)B_{0}(0, m_t^2, m_W^2)  \\
&+ 2 m_W^2 B_{0}(0,0,m_W^2)\big)\,,
\end{align}
\begin{align}
\nonumber \tilde{\mathcal{A}}_{(e)} =& \sin\theta\, \dfrac{\lambda_t\,m_t^2}{16\pi^2 (D-2) m_b v^3 (m_b^2-p_{h_{2}}^{2})}\cdot\\ 
&\big[2m_W^2(m_b^2-p_{h_{2}}^{2})B_{0}(0,m_W^2,m_W^2)\\\nonumber 
\nonumber &-(D-2)\big(m_b^4-m_b^2(m_t^2 + m_W^2 + 3 p_{h_{2}}^{2})  \\
\nonumber &+ p_{h_{2}}^{2}(m_t^2-m_W^2)\big) B_0(m_b^2, m_t^2, m_W^2)\big]\,,
\end{align}
\begin{align}
\nonumber \tilde{\mathcal{A}}_{(f)} = & -\sin\theta\,\dfrac{\lambda_t\, m_b}{8\pi^2 \,v^{3}(m_{b}^{2} - p_{h_{2}}^{2})}\,\Big\{m_{W}^{2} \big(2(2 - D) m_{W}^{2} \\\nonumber
 &+2 m_{b}^{2} - m_{t}^{2}\big) B_{0}(m_{b}^{2}, m_{t}^{2}, m_{W}^{2}) \\\nonumber
 &- 2 m_{W}^{2} \big(m_{b}^{2} - (D - 2) m_{W}^{2}\big) B_{0}(m_{b}^{2}, 0, m_{W}^{2}) \\\nonumber
 & + m_{t}^{2} (2 m_{W}^{2} + p_{h_{2}}^{2}) B_{0}(p_{h_{2}}^{2},  m_{W}^{2}, m_{W}^{2}) \\\nonumber
 &+ \big[m_{t}^{2} (2 m_{W}^{4} - m_{W}^{2} p_{h_{2}}^{2} + p_{h_{2}}^{4}) \\
\nonumber &- 4 m_{W}^{6} + 2  m_{b}^{4} m_{W}^{2} - m_{b}^{2} \big(m_{t}^{2} (2 m_{W}^{2} + p_{h_{2}}^{2}) \\\nonumber 
&+ 2 m_{W}^{2} p_{h_{2}}^{2}\big) + 
  m_{t}^{4} (2 m_{W}^{2} + p_{h_{2}}^{2})  \\\nonumber
  &+ 2 m_{W}^{4} p_{h_{2}}^{2}\big] C_{0}(0, m_{b}^{2} ,  p_{h_{2}}^{2},  m_{W}^{2},  m_{t}^{2}, m_{W}^{2}) \\ \nonumber
&- 2 m_{W}^{2} (-2 m_{W}^{4} + m_{b}^{4} - m_{b}^{2}  p_{h_{2}}^{2} \\
& + m_{W}^{2} p_{h_{2}}^{2}) C_{0}(0, m_{b}^{2} ,  p_{h_{2}}^{2},  m_{W}^{2}, 0, m_{W}^{2})\Big\}\,,
\end{align}
\begin{align}
\tilde{\mathcal{A}}_{(g)} &= -\sin\theta\, \frac{ \lambda_t m_t^4}{4\,\pi^2 (D-2) m_{b} v^3} B_{0}(0, m_t^2, m_{t}^{2})\,,
\end{align}
\begin{align}
\nonumber \tilde{\mathcal{A}}_{(h)} =&\sin\theta\,\dfrac{\lambda_{t}\,m_W^2}{8\,\pi^2 m_b v^3}\cdot  \\\nonumber
& \Big[ 
m_W^2(x-1)(D+x-2)B_{0}(0, m_t^2, m_W^2) \\\nonumber
& + \dfrac{2\,m_t^2}{D-2}\, B_{0}(0, m_W^2, m_W^2) \\ \nonumber
\nonumber & + (D-2)\, m_W^2 B_{0}(0,0, m_W^2) \\
&-2\, m_t^2 B_{0}(0, m_t^2, m_t^2)\Big]\,,
\end{align}
where $\lambda_t = V_{tb}\, V_{ts}^\ast$, $x=m_t^2/m_W^2$ and $D=4-2\epsilon$. The above results are to be multiplied with $\bar{s} P_R b$, where $s$ and $b$ denote the appropriate spinors and $P_R\equiv (1+\gamma_5)/2$.

Our definitions of Passarino-Veltman loop functions follow the \emph{Feyncalc}\ package \cite{Mertig:1990an,Shtabovenko:2016sxi,Shtabovenko:2020gxv}:
\begin{eqnarray}
&& \nonumber i\pi^2 B_{0}(p_{1}^{2}, m_{1}^{2} , m_{2}^{2}) \\
&&= \int  d^{D}k \frac{1}{(k^2 -  m_{1}^2) \left((k + p_{1})^2 -  m_{2}^{2}\right)}\, ,\\
\hspace{-15px}&& i\pi^2 C_{0}(p_{1}^{2}, (p_{1}-p_{2})^2,p_{2}^{2}, m_{1}^{2}, m_{2}^{2}, m_{3}^{2})  \\
&&  \nonumber  = \int d^{D}k \frac{1}{(k^2 -  m_{1}^2) \left((k + p_{1})^2 -  m_{2}^{2}\right) \left((k + p_{2})^2 -  m_{3}^{2} \right)}\,.
\end{eqnarray}

\section{Evaluation of the number of events at Belle
  II\label{App:Events}} 
We describe the formula for the evaluation of the number of events in
$B \to K^{(\ast)} h_2$, with the long-lived scalar $h_2$ decaying back
to $f$, a pair of leptons or hadrons at Belle II.

The energy and the magnitude of the momentum of $h_2$ in the $B$ meson
rest-frame are:
\begin{equation}
E_{h_2}=\dfrac{m_B^2+m_{h_{2}}^2-m_{K^{(\ast)}}^2}{2m_B}\,,\quad \vert\vec{p}_{h_{2}}\vert =\sqrt{E_{h_2}^2-m_{h_{2}}^2}\,.
\end{equation}
For our coordinate system we choose the $z$-axis in the direction of the electron beam. The convention for the angle $\vartheta$ follows Chapter 3 of Ref.~\cite{Kou:2018nap}. 
We consider the Lorentz transformation from the rest frame $h_2$ to the laboratory frame, $\mathcal{B}_1 \mathcal{R} \mathcal{B}_0 $, where $\mathcal{R} \mathcal{B}_0$ is the transformation from the rest frame of $h_2$ to the rest frame of the $B$ meson:
\begin{equation}
\qquad \mathcal{R}\mathcal{B}_0 \begin{pmatrix} m_{h_2}\\ 0\\ 0\\0 \end{pmatrix} = \begin{pmatrix} E_{h_2}\\ 0\\ \vert\vec{p}_{h_{2}}\vert \sin \vartheta_0 \\\vert\vec{p}_{h_{2}}\vert \cos \vartheta_0  \end{pmatrix}\,,
\end{equation} 
and $\mathcal{B}_1$ is the boost from the $\Upsilon$ rest frame to the laboratory frame. The $B$ meson pair is produced nearly at rest in the decay of the $\Upsilon$ resonance, so we neglect a small Lorentz boost from the $\Upsilon$ rest frame to the B rest frame. 
We also conveniently set the azimuthal angle $\phi$ to zero since it is not affected by the $\mathcal{B}_1$ boost along the $z$ direction. The latter boost is induced by the asymmetric beam energies $E_-= 7\,\text{GeV}$ and $E_+= 4\,\text{GeV}$ of electrons and positrons, respectively, and is determined by $\beta_B\gamma_B =(E_--E_+)/2 (E_- E_+)^{1/2} = 0.28$, $\gamma_B = 1.04$.

In the rest frame of the mediator, the decay occurs at $(c\tau, 0, 0, 0)$. The decay length in the laboratory frame follows from
\begin{align}
\quad \small
& 
\nonumber \begin{pmatrix} c t_{\text{lab}} \\ x_{\text{lab}} \\y_{\text{lab}}\\z_{\text{lab}}\end{pmatrix}\,= \mathcal{B}_1\mathcal{R}\mathcal{B}_0 \begin{pmatrix} c\tau\\ 0\\ 0\\0 \end{pmatrix} \\
&= \frac{c\tau}{m_{h_{2}}} \begin{pmatrix} \gamma_B & 0 & 0 & \gamma_B \beta_B\\  0 & 1 & 0 & 0\\0 & 0 & 1 & 0\\ \gamma_B \beta_B & 0 & 0 & \gamma_B \end{pmatrix} \begin{pmatrix}  E_{h_2}\\ 0\\ \vert\vec{p}_{h_{2}}\vert \sin \vartheta_0 \\ \vert\vec{p}_{h_{2}}\vert \cos \vartheta_0  \end{pmatrix} \\
\nonumber & =  \frac{c\tau}{m_{h_2}}\begin{pmatrix} \gamma_B E_{h_{2}} + \gamma_B \beta_B \vert\vec{p}_{h_{2}}\vert\cos\vartheta_0 \\ 0\\ \vert\vec{p}_{h_{2}}\vert \sin \vartheta_0 \\ \gamma_B \beta_B E_{h_{2}} +\gamma_B \vert\vec{p}_{h_{2}}\vert \cos \vartheta_0  \end{pmatrix}\,.
\end{align}

The decay length of the mediator in the laboratory frame is
$d_L=(x_{\text{lab}}^2 + y_{\text{lab}}^2 + z_{\text{lab}}^2)^{1/2}$ and
is related to the corresponding angle $\theta$ as
\begin{equation}
  y_{\text{lab}}= d_L(\theta_0)\sin\vartheta, \quad z_{\text{lab}}= d_L(\theta_0)\cos\vartheta\,.\label{Def:thetalab}
\end{equation}
The expected number of $B^\pm \to K^{(\ast)\pm} h_2[\to f] $ events
is
\begin{eqnarray}
\nonumber \quad N^{\pm}_{f}&=N_{B^+ B^-} \cdot 2\cdot B (B^\pm \to K^{(\ast)\pm}
                 h_2)
                 B (h_2\to f) \cdot \\
&\quad \int d\vartheta\,p(\vartheta)\frac{1}{d_L} \int_{r_{\text{min}}(\vartheta)}^{r_{\text{max}}(\vartheta)} dr e^{-\frac{r}{d_L}}\,,\label{Events-pm}
\end{eqnarray}
where $N_{B^+ B^-}$ is the total number of produced $B^+\text{-}B^-$
meson pairs. We include the differences in the lifetimes and the
production asymmetry of $B^+$ and $B^0$ mesons:
\begin{align}
\quad& \tau_{B^+} = 1.638\,\text{ps}\,,\qquad \tau_{B^0} = 1.519\,\text{ps}\,,\\
& f^{+-}\equiv B(\Upsilon(4S) \to B^+B^-) = 0.514\,,\\
     & f^{00}\equiv B(\Upsilon(4S) \to B^0\bar{B}^0) = 0.486\,,
       \label{bnums}
\end{align}
where the numerical values are taken from \cite{Amhis:2019ckw}.
The total number of the displaced vertex events, summed
over the decays of $B^+, B^-, B^0$ and $\bar{B}^0$ mesons, is
\begin{eqnarray}
  \nonumber\quad   N^{\text{tot}}_{f} &=&  N_{B \bar{B}} \cdot 2\cdot
              B (B^\pm \to K^{(\ast)\pm}
              h_2) B (h_2\to f)\cdot \\
                  & & \quad \big(f^{+-} +
                    f^{00}\frac{\tau_{B^0}}{\tau_{B^+}}\big)\cdot \nonumber\\
&&\quad \int d\vartheta\,p(\vartheta)\frac{1}{d_L}
          \int_{r_{\text{min}}(\vartheta)}^{r_{\text{max}}(\vartheta)} dr e^{-\frac{r}{d_L}} \,,
\label{Events-corr}
\end{eqnarray}
where
$N_{B\bar{B}}\equiv N_{B^+ B^-}+N_{B^0 \bar{B}^0} = 5\cdot 10^{10}$ is
the total number of produced $B$ meson pairs with $50\,\text{ab}^{-1}$
of data at the Belle II experiment \cite{Kou:2018nap}.  
With \eq{bnums} we find
\begin{eqnarray}
  \nonumber \quad N^{\text{tot}}_{f} &=& N_{B \bar{B}} \cdot 1.93\cdot B (B^\pm \to K^{(\ast)\pm}
                                   h_2) B (h_2\to f) \cdot \\
      &&\quad  \int d\vartheta\,p(\vartheta)\frac{1
                                  }{d_L}
  \int_{r_{\text{min}}(\vartheta)}^{r_{\text{max}}(\vartheta)} dr e^{-\frac{r}{d_L}}\,.\label{Events}
\end{eqnarray}
The angular distribution of the mediator in the $B$ meson rest frame is trivial:  
\begin{equation}
\qquad p(\vartheta_0)=\frac{1}{2}\sin\vartheta_0\,,
\end{equation}
whereas the distribution with respect to the angle in the laboratory frame $\vartheta$ is
\begin{equation}
  \qquad p(\vartheta) =
  \frac{1}{2}\sin\vartheta_0 \bigg\vert \frac{d\vartheta_0}{d\vartheta}\bigg\vert\,,
\end{equation}
where we can express the angle $\vartheta_0$ in terms of $\vartheta$
using eq.~\eqref{Def:thetalab}.

The maximally travelled distance in the Belle II detector as a function
of the angle $\vartheta$ is given by the geometry of the compact drift
chamber (CDC). Following Chapter 3 of Ref.~\cite{Kou:2018nap} we find:
\begin{align}
\nonumber \quad& \vartheta\in (0.3, \arctan \frac{h}{d_1})\,,\quad r_{\text{max}} = \frac{d_1}{\cos\vartheta}\,,\\ 
& \vartheta\in (\arctan \frac{h}{d_1}, \frac{\pi}{2} + \arctan
   \frac{d_2}{h})\,,\quad r_{\text{max}} = \frac{h}{\sin\vartheta}\,,\nonumber \\
& \vartheta\in (\frac{\pi}{2} + \arctan \frac{d_2}{h}, \frac{5\pi}{6})\,,\quad r_{\text{max}} = -\frac{d_2}{\cos\vartheta}\,,
\end{align}
where $d_1$ ($d_2$) is the dimension of the CDC along the positive
(negative) $z$-direction measured from the interaction point and $h$ is
the height measured from the beam line. In our evaluation we use
$d_1=1.5\,$m, $d_2=0.74$m, $h=1.17\,$m.

Following Ref.~\cite{Filimonova:2019tuy} we use for the minimal vertex
resolution $r_{\text{min}}=500\,\mu m$ in the formula \eqref{Events},
but neglect its dependence on $\vartheta$. Our final formula is:
\begin{align}
  N^{\text{tot}}_{f} &=N_{B \bar{B}} \cdot 1.93\cdot B (B^\pm \to K^{(\ast)\pm} h_2)
          B (h_2\to f)
          \cdot \nonumber \\
   &\quad
              \int d\vartheta\,\sin\vartheta_0(\vartheta) \bigg\vert
              \frac{d\vartheta_0(\vartheta)}{d\vartheta}\bigg\vert\,\Big(e^{-\frac{r_{\text{min}}}{d_L(\vartheta)}}
              -
              e^{-\frac{r_{\text{max}}(\vartheta)}{d_L(\vartheta)}}\Big)\,. \label{Events2} 
\end{align}
We tabulate the total number of displaced vertex events $N^{\text{tot}}_f$ for
interesting values for the proper lifetime $\tau$
and mass of $h_2$ in Tab.~\ref{tab:1}. 

\begin{table}[!ht]
  \hrule ~\\[2mm]
  \centerline{\footnotesize{
    \begin{tabular}{*{6}{|c}|}\hline
      \multirow{2}{*}{$m_{h_2}\text{[GeV]}$} & \multicolumn{5}{|c|}{$\tau\,[\text{ps}]$} \\ \cline{2-6}
       & $250$ & $500$ & $1000$ &$2000$ &$4000$ \\ \hline
       $0.3$ & 50204 & 18385 & 5734 & 1614 & 429\\ 
       $0.9$ & 972.3 & 465 & 191.8 & 65.7 & 19.6\\ 
       $1.5$ & 1634.7 & 815.2 & 382.7 & 152.7 & 50.9\\ 
       $2.1$ & 334.2 & 167.6 & 82.6 & 36.8 & 13.7\\ 
       $2.7$ & 115.6 & 58 & 29 & 13.9 & 5.8\\ 
       $3.3$ & 56.8 & 28.6 & 14.4 & 7.1 & 3.2\\ 
       $3.9$ & 58.4 & 29.6 & 14.9 & 7.4 & 3.6\\ \hline
    \end{tabular}}}
  \caption{Total number $N^{\text{tot}}_f$ of displaced-vertex
    $B \to K^{(\ast)} h_2[\to f] $ events (see
    \eqref{Events2}) occuring in the CDC of Belle II
    for various values of the proper lifetime
    (columns) and mass (rows) of $h_2$. All charges of the
    final state mesons $K$ and $K^{\ast}$ are included, as well as
    $f=\mu\mu, \tau\tau, \pi\pi, KK$.}\label{tab:1}
~\\[-3mm]\hrule  
 \end{table}


\end{document}